\documentclass[11pt]{article}
\usepackage{amssymb, amsmath, amsthm, amsfonts, amscd, epsfig}
\usepackage[english]{babel}
\usepackage{bm}
\usepackage{graphicx, epsfig,subcaption}
\usepackage{color}
\usepackage{placeins}
\usepackage{enumerate}

\def\ep{\varepsilon}

\newcommand{\Om}{\Omega}

\newcommand{\RR}{\mathbb{R}}
\newcommand{\CC}{\mathbb{C}}

\newcommand{\p}{\partial}

\newcommand{\ds}{\displaystyle}
\newcommand{\eqnref}[1]{(\ref {#1})}

\newcommand{\beq}{\begin{equation}}
\newcommand{\eeq}{\end{equation}}

\usepackage{setspace}

\numberwithin{equation}{section}
\numberwithin{figure}{section}

\begin{document}

\title{Shielding at a distance due to anomalous resonance
\thanks{\footnotesize
This work is supported by the Korean Ministry of Science, ICT and Future Planning through NRF grant Nos. 2013R1A1A3012931 (to M.L) and 2016R1A2B4014530 (to M.L).}}

\author{
Sanghyeon Yu\thanks{ Department of Mathematics, ETH Z\"urich, R\"amistrasse 101, CH-8092 Z\"urich, Switzerland (sanghyeon.yu@sam.math.ethz.ch).}
 \and Mikyoung
Lim
\thanks{Department of Mathematical Sciences,
Korea Advanced Institute of Science and Technology, Daejeon
305-701, Korea (mklim@kaist.ac.kr).}
}

\maketitle

\begin{abstract}
{\color{black}A cylindrical plasmonic structure with a concentric core exhibits an anomalous localized resonance which results in cloaking effects. Here we show that, if the structure has an eccentric core, a new kind of shielding effect can happen.
In contrast to the conventional shielding device, our proposed structure can block the effect of external electrical sources even on a region which is not enclosed by any conducting materials. In fact, the shielded region is located at a distance from the device.
We analytically investigate this phenomenon by using the M\"{o}bius 
transformation via which an eccentric annulus is transformed into a concentric one. We also present several numerical examples.}
\end{abstract}
\section{Introduction}

A cylindrical superlens with a shell having the relative permittivity $\ep_\delta=-1+i\delta$ exhibits an anomalous resonant behavior as the loss parameter $\delta$ tends to zero \cite{NMM94,MNMP_PRSA_05}. 
{The electric field distribution generated by this structure diverges in magnitude throughout a region localized within a specific distance from the superlens, while it converges to a smooth field farther away from the superlens.} 
Furthermore, the localized region, which changes depending on the source location, has a boundary that does not coincide with any discontinuity in the permittivity distribution. 
This so-called anomalous localized resonance was first discovered by Nicorovici, McPhedran and Milton \cite{NMM94} and is responsible for the subwavelength resolving power of the superlens \cite{Pendry03}; see also \cite{Veselago, Pendry00}. 
The superlens acts as a cloaking device for certain sources since the resonance cancels the effect of those sources \cite{MN06}. 
Extensive work has been produced on cloaking by anomalous localized resonance, for example in \cite{ACKLM13,ACKLM13_PRSLS,KSW14_CMP, MNMP_PRSA_05, NMM94,Nguyen15_JEMS,Nguyen15}.

In this paper, we show that a cylindrical superlens can also act as a new kind of {quasistatic}  shielding device if the core is eccentric to the shell. The historical root of {the shielding effect}  reaches back to 1896 when Michael Faraday discovered that a region coated with a conducting material is not affected by external electric fields. 
While such a conventional method shields a region enclosed by the device, a superlens with an eccentric core can shield a non-coated region which is located outside the device. 
We call this phenomenon {\it shielding at a distance}. The aim of this paper is to investigate the conditions required for shielding at a distance and geometric features such as the location and size of the shielded region.
{\color{black} The key element to study in the eccentric case is the M\"{o}bius transformation via which a concentric annulus is transformed into an eccentric one.
The {quasistatic} properties of the eccentric superlens can be derived in a straightforward way from those of the concentric case since the M\"{o}bius transformation is a conformal mapping.}

\begin{figure*}
   \centering 
      \centering
     \begin{subfigure}{0.35\textwidth}
      \centering
      \includegraphics[height=5.0cm]{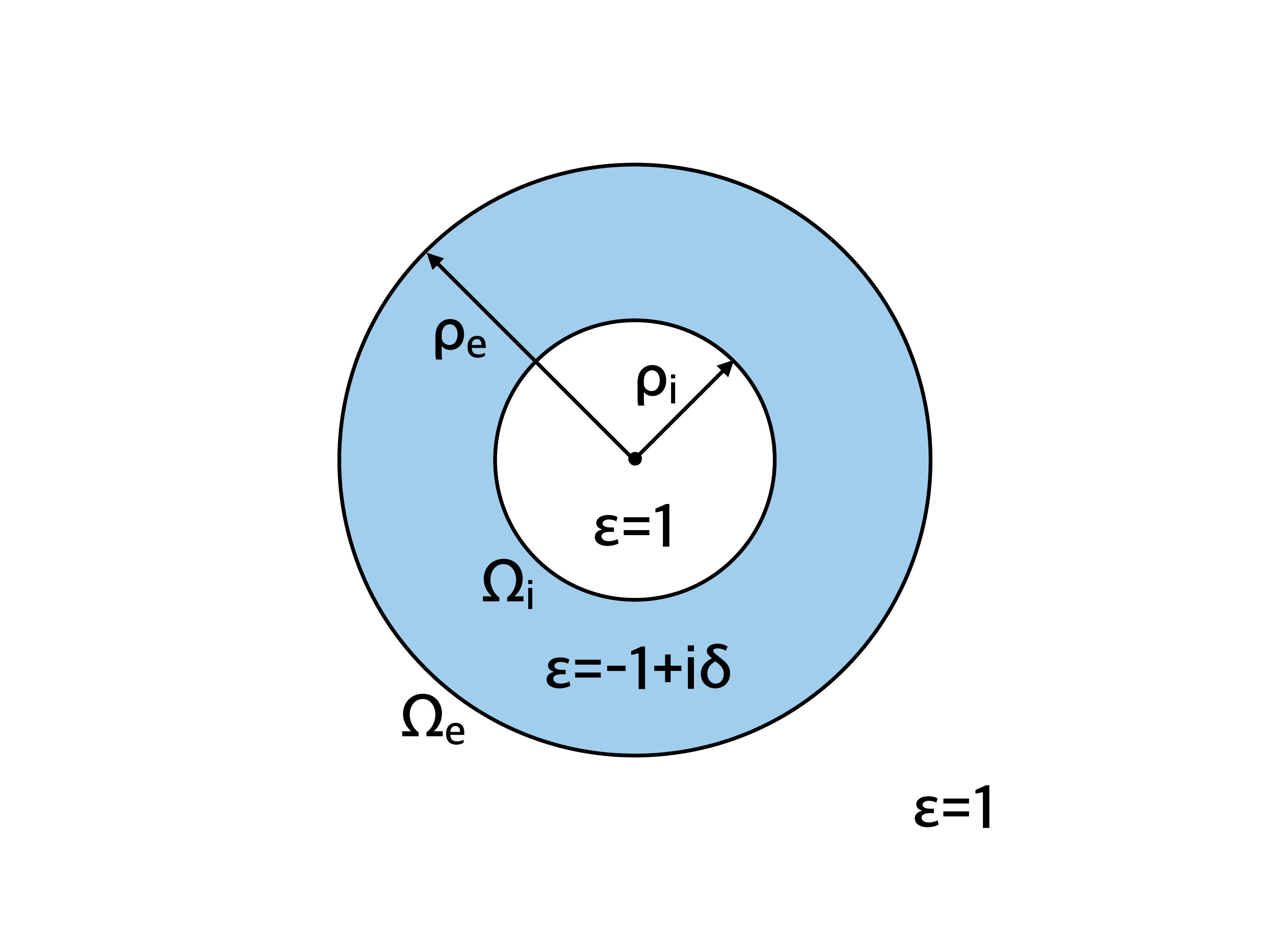}
      \caption{}
    \end{subfigure}%
    \hskip1cm
     \begin{subfigure}{0.35\textwidth}
      \centering
      \includegraphics[height=5.0cm]{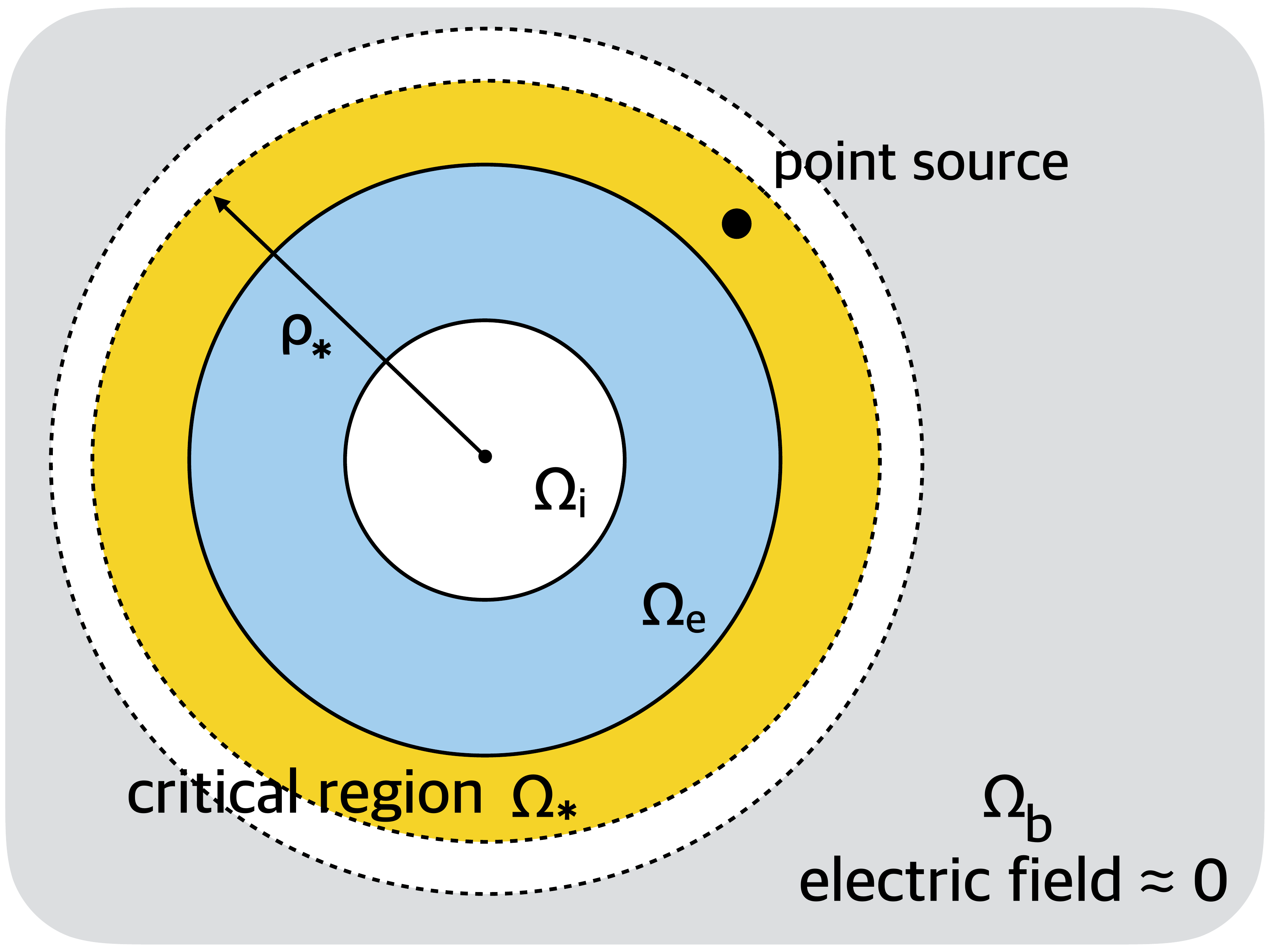}
      \caption{}
      \end{subfigure}
    \caption{{\color{black}Cloaking due to the anomalous localized resonance: (a) shows the structure of the superlens with concentric core; (b) illustrates the cloaking effect.}}
    \label{superlens_con}
    \end{figure*}

{
The anomalous resonance in the eccentric case can be very different from that of the concentric case. Interestingly, there can be an undisturbed region of the electric fields inside a resonance region.
Also, the resonance can happen even for the electric sources at infinity.
 Our proposed shielding device is based on these two features of the anomalous resonance.
We emphasize that this unusual resonance might have more important applications besides the shielding.

Throughout this paper, the superlens is assumed to be much smaller than operating wavelength and, thus, the quasistatic approximation is valid. While we assume the smallness for the superlens, we do not restrict the range of location for the external sources.
Although the concentric superlens cause the cloaking only for the sources well located within a wavelength, the eccentric one can induce the shielding even for the sources located far away from the structure. 
This feature will be explained in a later section.


}

The paper is organized as follows. In section \ref{sec:ALR_concentric}, we review cloaking due to the anomalous localized resonance {\color{black} caused by the concentric superlens.}
In section \ref{sec:Mobius}, we introduce the M\"{o}bius transformation and discuss its geometrical properties. In section \ref{sec:potential}, we consider how the electric potential transforms by the M\"{o}bius transformation. In section \ref{sec:shielding}, we explain how a new shielding effect can happen and derive the explicit conditions required for its occurrence. Finally, we illustrate shielding at a distance by presenting several numerical examples in section \ref{sec:numerical}.%

\section{Anomalous localized resonance {\color{black} caused by the concentric superlens}}\label{sec:ALR_concentric}

{\color{black}
%
%
%

In this section, we review the anomalous resonance caused by the concentric superlens whose geometry is described in Figure \ref{superlens_con}(a). 

We first fix some notations to explicitly state the phenomenon.  
We let $\Om_i$ and $\Om_e$ denote circular disks centered at the origin with the radii $\rho_i$ and $\rho_e$, respectively, satisfying $0<\rho_i<\rho_e<1$. 
Identifying $\mathbb{R}^2$ as $\mathbb{C}$, they can be represented as
$$
{\Omega}_i=\big\{z\in\mathbb{C}:|z|<\rho_i\big\}\quad\mbox{and}\quad {\Omega}_e=\big\{z\in\mathbb{C}:|z|<\rho_e\big\}.
$$
The core $\Omega_i$ and the background $\RR^2\setminus\overline{\Omega_e}$ are assumed to be occupied by the isotropic material of permittivity $1$ and the shell $\Omega_e\setminus\overline{\Omega_i}$ by the plasmonic material of permittivity $-1+i\delta$ with a given loss parameter $\delta>0$, {\it i.e.}, 
the permittivity distribution $\ep_\delta$ is given by
\beq\label{epdelta_new}
{\ep}_\delta=
\begin{cases}
1\quad&\mbox{in the core},\\
-1+i\delta\quad&\mbox{in the shell},\\
1\quad&\mbox{in the background}.
\end{cases}
\eeq
}
%
We also assume the annulus structure to be small compared to the operating wavelength so that it can adopt the quasistatic approximation. Then the (quasistatic) electric potential ${V}_\delta$ satisfies
\beq\label{eqn:cond_new}
\nabla\cdot{\ep}_\delta\nabla {V}_\delta = {f}\quad\mbox{in }\mathbb{C},
\eeq
where ${f}$ represents an electrical source. We assume that $f$ is a point multipole source of order $n$ located at a location $z_0\in\RR^2\setminus\overline{\Omega_e}$. Then the potential $ F$ generated by the source $ f$ can be represented as
$$
 F(z)=\sum_{k=1}^n\mbox{Re}\{ c_k (z-z_0)^{-k}\}, \quad z\in\CC,
$$ 
with complex coefficients $c_k$'s.  When $n=1$, the source $ f$ (or the potential $ F$) means a point dipole source.

Now we discuss the anomalous localized resonance.
 Its rigorous mathematical description was given in \cite{ACKLM13} as follows:
\begin{itemize}
\item[(i)]
 the dissipation energy ${W}_\delta$ diverges as the loss parameter $\delta$ goes to zero if and only if a point source $ f$ is located inside the region ${\Omega}_*:=\{|z|<\rho_* \}$, where $\rho_*:=\sqrt{\rho_e^3/\rho_i}$ and ${W}_\delta$ is given by
\beq
\label{ALR}{W}_\delta:=\mbox{Im} \int_{\RR^2}\ep_\delta|\nabla V_\delta|^2\;dx=\delta\int_{\Omega_e\setminus\overline{\Omega_i}} |\nabla {V}_\delta|^2.
\eeq 
Let us call ${\Omega}_*$ (or $\rho_*$) {\it the critical region} (or {\it the critical radius}), respectively.
\item[(ii)] 
the electric field $-\nabla V_\delta$ stays bounded outside some circular region regardless of $\delta$.  More precisely, we have
\beq
|\nabla V_\delta(z)| \leq C, \quad z \in {\Omega_b}:=\{|z|>\rho_e^2/\rho_i\},
\eeq
for some constant $C$ independent of $\delta$. Here, the subscript `b' in $\Omega_b$ indicates the boundedness of the electric field. Let us call $\Omega_b$ {\it the calm region}.

\end{itemize}

Now we explain why the above statements represent the cloaking effect. Suppose that the point source $f$ is located inside the critical region ${\Omega}_*$. Then, by the fact (i), the energy $W_\delta$ goes to infinity as $\delta\rightarrow 0$.
Since the blow-up of the energy ${W}_\delta$ is unphysical, we have to  consider the normalized potential ${V}_\delta/\sqrt{W_\delta}$  instead of $V_\delta$. The corresponding energy will remain bounded regardless of $\delta$. Then, by the fact (ii), the normalized electric field $-\nabla V_\delta/\sqrt{W_\delta}$ goes to zero as $\delta\rightarrow 0$ in the region $\Omega_b$. In other words, assuming the smallness of $\delta$, the field generated by the point source $f$ is negligible, which means that the point source $f$ is cloaked; see Figure \ref{superlens_con}(b). In contrast, if the source $f$ is located outside the critical region $\Omega_*$, then the energy dissipation does not blow up. We can detect the presence of the source by observing the potential away from the superlens since the electric field is not negligible on $\Omega_b$ everywhere.

\section{M\"obius transformation}\label{sec:Mobius}
In this section, we will show that
the concentric annulus can be transformed into an eccentric one by applying the M\"{o}bius transformation $\Phi$ defined as
\beq\label{Phi_def_new}
\zeta=\Phi(z) := a\frac{z+1}{z-1}
\eeq
with a given positive number $a$. 
We shall also discuss how the critical region is transformed depending on the ciritical paramter $\rho_*$.

\begin{figure*}
\begin{center}
\hskip 1mm
  \includegraphics[width=7.0cm]{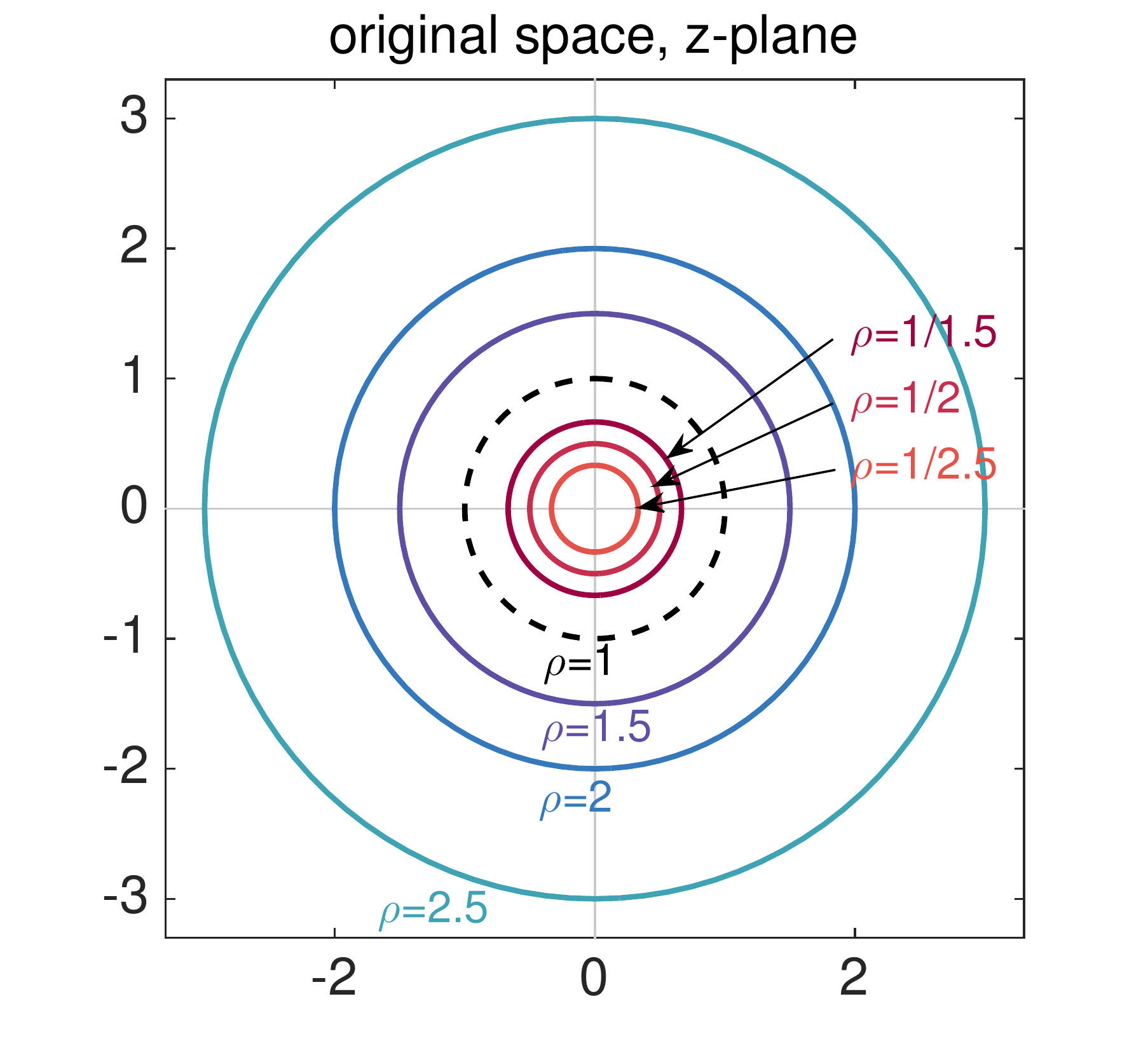}\hskip 1mm
\includegraphics[width=7.0cm]{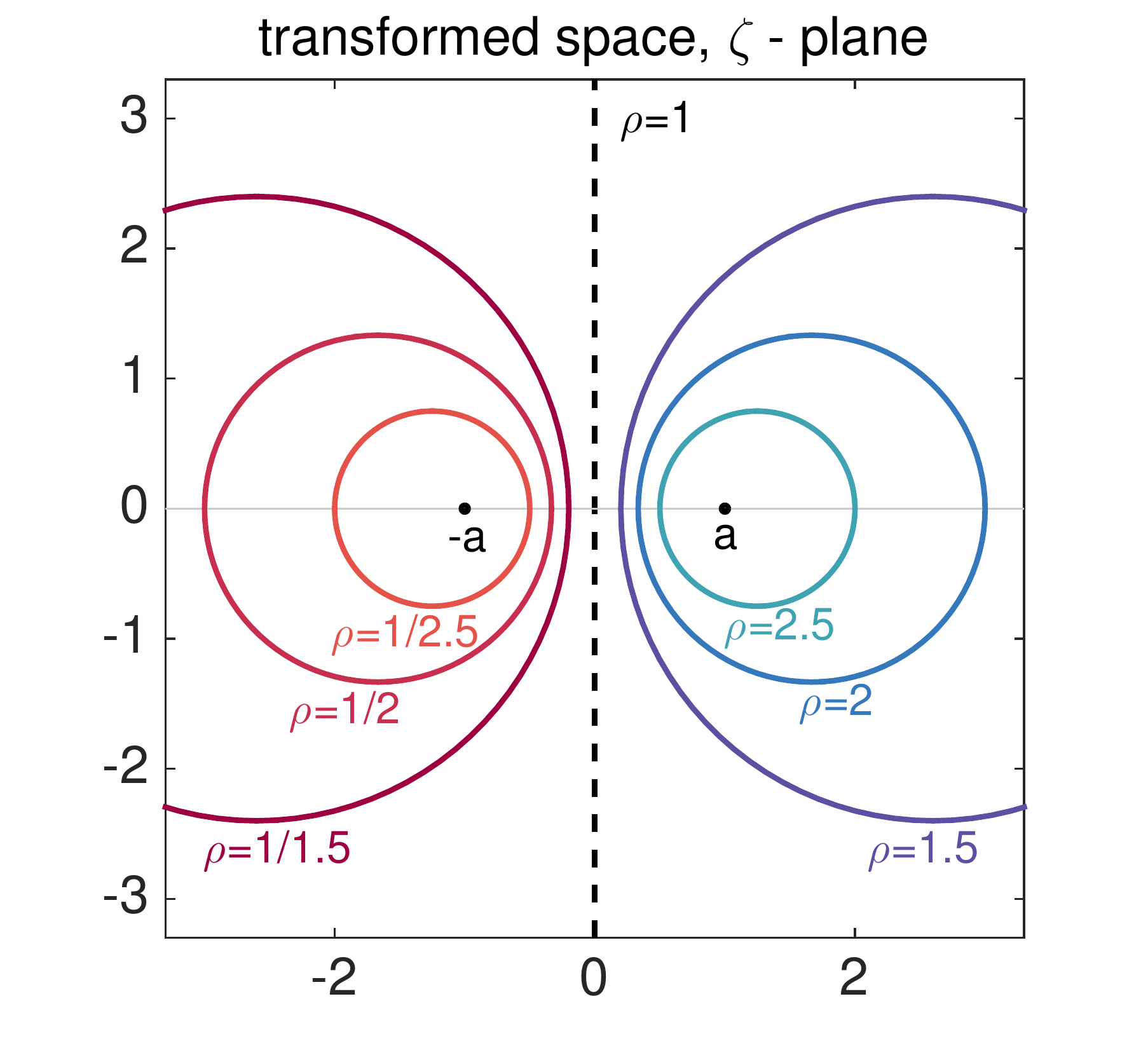}\hskip 1mm
\end{center}
\caption{{\color{black}The  M\"{o}bius transformation $\Phi$ defined in \eqnref{Phi_def_new} maps  $0$, $\infty$, $1$ to $-a$, $+a$, $\infty$, respectively. The left figure shows radial coordinate curves $\{|z|=\rho\}$, $\rho>0$, and the right figure their images transformed by $\Phi$ with $a=1$.  Concentric circles satisfying $\rho\neq1$ are transformed into eccentric ones.}}
\label{Phi_fig}
\end{figure*}

The function $\Phi$ is a conformal mapping from $\mathbb{C}\setminus\{1\}$ to $\mathbb{C}\setminus\{a\}$. It maps the point $z=1$ to infinity, infinity to $\zeta=a$, and $z=0$ to $\zeta=-a$. It maps a circle centered at the origin, say $S_\rho:=\{z\in\mathbb{C}:|z|=\rho\}$, to the circle given by
\beq\label{circ_to_circ}
 \Phi(S_\rho)=
  \ds\{z\in\mathbb{C}: |z-c| = r \}, 
 \quad\mbox{where }  c=a\frac{\rho^2+1}{\rho^2-1} \mbox{ and } r=\frac{2a}{|\rho-\rho^{-1}|}. 
\eeq
So the concentric circles $S_\rho$'s with $\rho\neq1$ are transformed to eccentric ones in $\zeta$-plane; see Figure \ref{Phi_fig}.

{\color{black} Let us discuss how the concentric superlens described in section \ref{sec:ALR_concentric} is geometrically transformed by the mapping $\Phi$.
Note that for $0<\rho<1$, the transformed circle $\Phi(S_\rho)$ always lies in the left half-plane of $\mathbb{C}$. Since we assume that $0<\rho_i<\rho_e<1$, the concentric annulus in $z$-plane is changed to an eccentric one contained in the left half $\zeta$-plane. We let $\widetilde{\Omega}_i$ (or $\widetilde \Omega_e$) denote the transformed disk of $\Om_i$ (or $\Om_e$), respectively.
}

Now we consider the critical region $\Omega_*=\{|z|<\rho_*\}$ and the calm region ${\Omega}_b$. Let us denote the transformed critical region (or calm region) by $\widetilde{\Omega}_*$ (or $\widetilde{\Omega}_b$), respectively.
The shape of  $\widetilde \Omega_*$  can be very different depending on the value of $\rho_*$.  Suppose $0<\rho_*<1$ for a moment. Then the  region $\widetilde \Omega_*$ is a circular disk contained in the left half $\zeta$-plane. Next, assume that $\rho_*>1$. In this case, $\widetilde \Omega_*$ becomes the region outside a disk which is disjoint from the eccentric annulus. Contrary to the case when $\rho_*<1$, the region $\widetilde\Omega_*$ is now unbounded.
Similarly, the shape of $\widetilde \Omega_b$ depends on the paramter $\rho_b:=\rho_e^2/\rho_i$. If $0<\rho_b<1$, $\widetilde \Omega_b$ is a region outside a circle. But, if $\rho_b>1$, $\widetilde \Omega_b$ becomes a bounded circular region which does not intersect with the eccentric superlens.
This unbounded (or bounded) feature of the shape of $\widetilde \Omega_*$ (or $\widetilde \Omega_b$) will be essentially used to design a new shielding device. 





\section{Potential in the transformed space}\label{sec:potential}

{
Here, we will transform the potential $V_\delta$ via the M\"{o}bius map $\Phi$ and then show that the resulting potential describes the physics of the eccentric superlens.
Let us define the transformed potential $\widetilde{V}_\delta$ by
 $\widetilde{V}_\delta(\zeta):={V}_\delta\circ\Phi^{-1}(\zeta)$.
} 
Since the M\"{o}bius transformation $\Phi$ is a conformal mapping, it preserves the harmonicity of the potential and interface conditions. It can be easily shown that the transformed potential ${V}_\delta$ satisfies
\beq
\ds\nabla\cdot \widetilde{\ep}_\delta\nabla \widetilde{V}_\delta=\widetilde{f}\quad\mbox{in }\mathbb{C},
\eeq
where $\widetilde{f}(\zeta)=\frac{1}{|\Phi'|^2}{(f\circ \Phi^{-1})}(\zeta)$ and the permittivity $\widetilde{\ep}_\delta$ is given by 
\beq\label{eccentric:material}
\widetilde{\ep}_\delta(\zeta)=
\begin{cases}
1\quad&\mbox{in }\widetilde{\Omega}_i,\\
-1+i\delta \quad &\mbox{in } \widetilde\Omega_e\setminus \overline{\widetilde\Omega_i},\\
1 \quad &\mbox{in the background}.
\end{cases}
\eeq
{\color{black}
    Therefore, the transformed potential $\widetilde{V}_\delta$ represents the quasistatic electrical potential of the eccentric superlens \eqnref{eccentric:material} induced by the source $\tilde{f}(\zeta)$.} 

Now we consider some physical properties in the transformed space.
The dissipation energy $\widetilde{W}_\delta$ in the transformed space turns out to be the same as the original one ${W}_\delta$ as follows:
\begin{align}
\widetilde{W}_\delta&
=\delta \int_{\p(\widetilde{\Omega}_e\setminus\widetilde{\Omega}_i)}  \widetilde{V}_\delta \frac{\p \widetilde{V}_\delta}{\p\widetilde{n}}  \; d\widetilde{l}
= \delta \int_{\p (\Omega_e\setminus\Omega_i)}V_\delta \frac{1}{|\Phi'|}\frac{\p V_\delta}{\p{n}} |\Phi'|\; dl
 = W_\delta.
 \end{align}
In the derviation we have used the Green's identity and the harmonicity of the potentials $V_\delta$ and $\widetilde{V}_\delta$.

The point source $f$ is transformed into another point source at a different location. To see this, we
recall that the source $f$ is located at $z=z_0$  in the original space. It
generates the potential  $F(z)=\sum_{k=1}^n\mbox{Re}\{ c_k (z-z_0)^{-k}\}$.  By the map $\Phi$, the potential $F$ becomes $\widetilde{F}:=F\circ \Phi^{-1}$ which is of the following form:
\beq
\widetilde{F}(\zeta)=\sum_{k=1}^n \mbox{Re}\left\{d_n (\zeta-\zeta_0)^{-k}\right\},\eeq
where $d_k$'s are complex constants and $\zeta_0:=\Phi(z_0)$. So the transformed source $\widetilde{f}$ is a point multipole source of order $n$ located at $\zeta=\zeta_0$.
It is also worth remarking that, if the point source $f$ is located at $z_0=1$ in the originial space, then $\widetilde{f}$ becomes a multipole source at infinity in the transformed space. In fact, its corresponding potential $\widetilde{F}$ is of the following form:
 $$\widetilde{F}(\zeta)=\sum_{k=1}^n \mbox{Re}\left\{e_k\zeta^k\right\}$$
for some complex constants $e_k$. For example, if $n=1$, then the source $\widetilde{f}$ (or potential $\widetilde{F}$) represents a uniform incident field.

\section{Shielding at a distance due to anomalous resonance}\label{sec:shielding}

In this section, we analyze the anomalous resonance in the eccentric annulus and explain how a new kind of shielding effect can arise.
In view of the previous section, the mathematical description of anomalous resonance in the eccentric case can be directly obtained from that in the concentric case as follows:
\begin{itemize}
\item[(i)]
 the dissipation energy $\widetilde{W}_\delta$ diverges as the loss parameter $\delta$ goes to zero if and only if a point source $\widetilde f$ is located inside the region $\widetilde{\Omega}_*$.
\item[(ii)] the electric field $-\nabla \widetilde{V}_\delta$ stays bounded  in the calm region $\widetilde{\Omega}_b$ regardless of $\delta$, {\it i.e.},
\beq
|\nabla \widetilde{V}_\delta(\zeta)| \leq C, \quad \zeta \in  { \widetilde{\Omega}_b},
\eeq
for some constant $C$ independent of $\delta$.
\end{itemize}

Now we discuss a new shielding effect. Suppose the parameters $\rho_i$ and $\rho_e$ satisfy $\rho_*=\sqrt{\rho_e^3/\rho_i}>1$.    
Then, as explained in section \ref{sec:Mobius}, the calm region $\widetilde \Omega_b$ becomes a bounded circular region which does not intersect with the eccentric structure.
If a point source is located within the critical region $\widetilde{\Omega}_*$, then the anomalous resonance occurs and the normalized electric field $-\nabla V_\delta/\sqrt{E}_\delta$ is nearly zero inside the calm region $\widetilde \Omega_b$.
So the bounded circular region $\widetilde \Omega_b$ is not affected by any surrounding  point source {\color{black} located in $\widetilde{\Om}_*$}. In other words, the shielding effect does occur in $\widetilde \Omega_b$, but there is a significant difference in this shielding effect compared to the standard one. There is no additional material enclosing the region $\widetilde \Omega_b$; the eccentric structure is located disjointly. So we call this effect `shielding at a distance' and $\widetilde \Omega_b$ `the shielding region'. The condition for its occurrence can be summarized as follows:
shielding at a distance happens in $\widetilde \Omega_b$ if and only if the critical parameter $\rho_*$ and the source location $\zeta_0$ satisfy
\beq\label{shield_criteria}
\rho_* > 1 \mbox{ and } \zeta_0 \in {\widetilde \Omega_*}.
\eeq


\begin{figure*}
   \centering 
     \begin{subfigure}{0.35\textwidth}
      \centering
      \includegraphics[height=5.0cm]{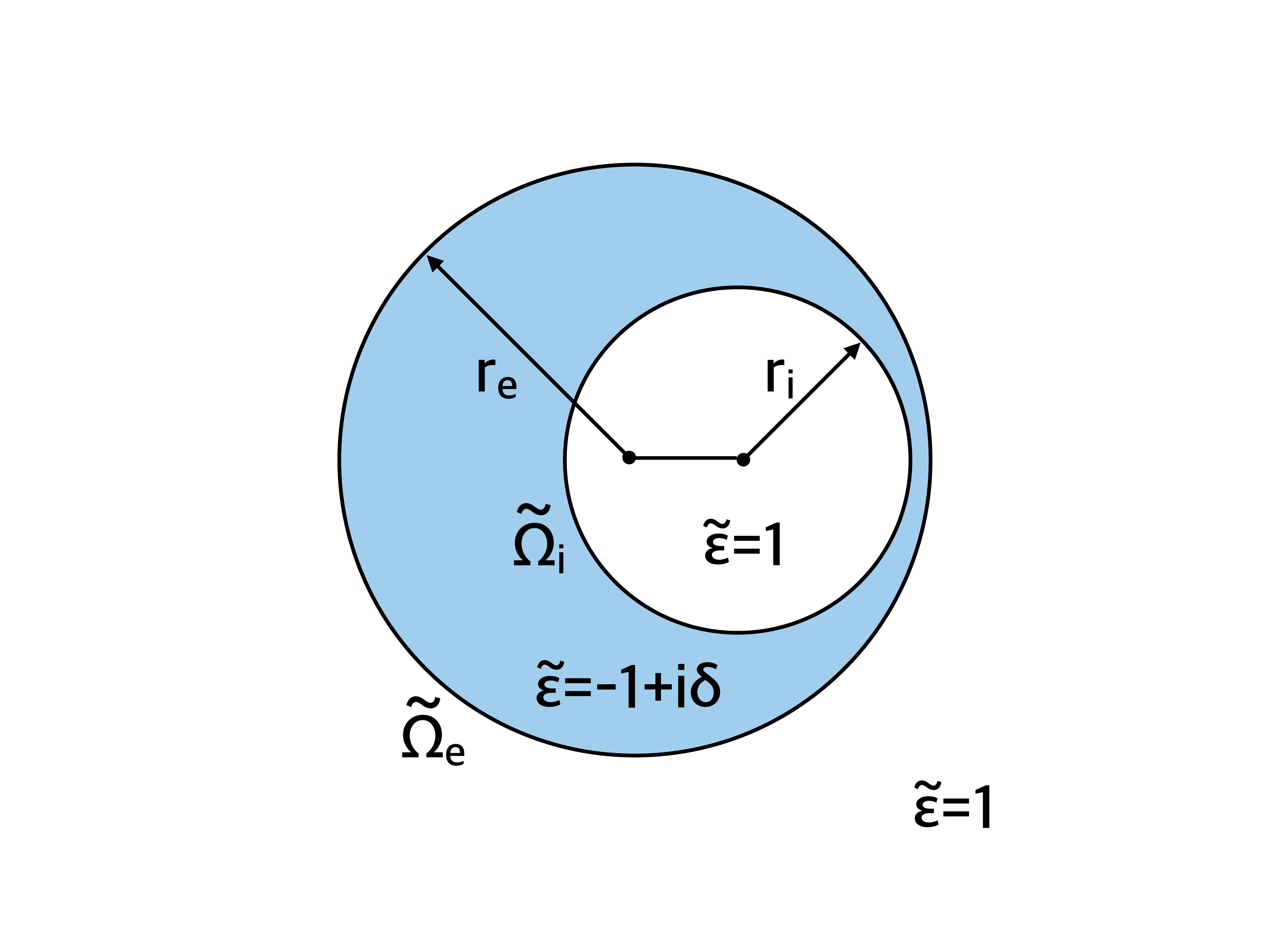}
      \caption{}
    \end{subfigure}%
    \hskip1cm
     \begin{subfigure}{0.35\textwidth}
      \centering
      \includegraphics[height=5.0cm]{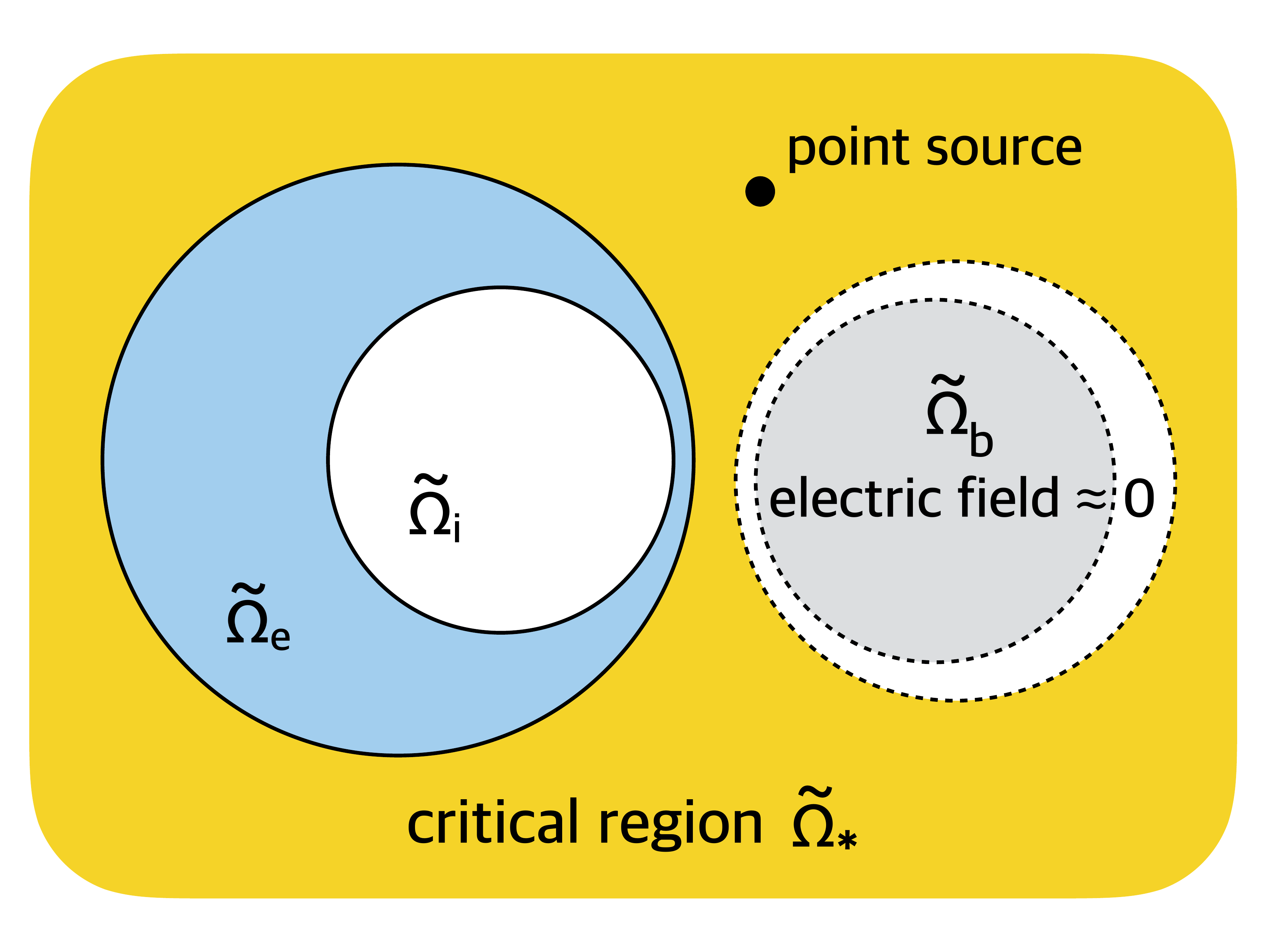}
      \caption{}
       \end{subfigure}%
    \caption{{\color{black} Shielding at a distance due to the anomalous localized resonance: (a) shows the structure of the superlens with the eccentric core; (b) illustrates shielding at a distance.} }
    \label{superlens}
    \end{figure*}

 The shielding effect occurs for not only a point source but also an external field like a uniform incident field {\color{black}$\widetilde F(\zeta)=- \mbox{Re}\{E_0\zeta\}$ for a complex constant $E_0$. As mentioned previously, an external field of the form $\mbox{Re}\{\sum_{k=1}^n e_k \zeta^k\}$ can be considered as a point source at $\zeta=\infty$.} Since the critical region $\widetilde \Omega_*$ contains the point at inifinity when $\rho_*>1$, the anomalous resonance will happen and then the circular bounded region $\widetilde \Omega_b$ will be shielded. {\color{black}It is worth remarking that, unlike in the eccentric case, the anomalous resonance cannot result from any external field with source at infinity for the concentric case.}
 


{
We now discuss the range of validity of the new shielding effect. 
Recall that we assume the eccentric superlens has a negative permittivity. For this, we may use plasmonic materials (such as gold and silver) whose permittivities have negative real part in the infrared to visible regime. 
In these regimes, the operating wavelength lies in the range of several hundreds of nanometers. 
Then the eccentric structure and the shielded region should have less than a few tens of nanometers in size because our analysis is based on the quasistatic approximation. Also, they should be relatively close to each other.
We, however, emphasize that the sources don't have to be located near the eccentric structure.
In fact, the shielding can happen for external electromagnetic waves surrounding the structure and the shielded region. Suppose, for instance, that a plane wave is incident. Since the eccentric structure is small compared to the wavelength, the incident field is nearly uniform across the structure. As already explained, the anomalous resonance can happen for the uniform field, which is a point dipole at infinity. So the shielding at a distance will occur for the plane wave. Similarly, it can happen for general surrounding waves.



}

\section{Numerical illustration}\label{sec:numerical}
In this section we illustrate shielding at a distance by showing several examples of the field distribution generated by an eccentric annulus and a point source. To compute the field distribution, we use an analytic solution derived by applying a  separation of variables method in the polar coordinates to the concentric case and then using the M\"{o}bius transformation $\Phi$. Although we omit the details, we refer to \cite{MN06, ACKLM13} for the analytic solution in the concentric case.

For all the examples below, we fix $\rho_e=0.7$ for the concentric shell and $a=1$ for the M\"{o}bius transformation. We also fix the loss parameter as $\delta = 10^{-12}$.

\subsection{Cloaking of a dipole source} \label{subsection:cloaking}
\begin{figure*}
\begin{center}
{\epsfig{figure=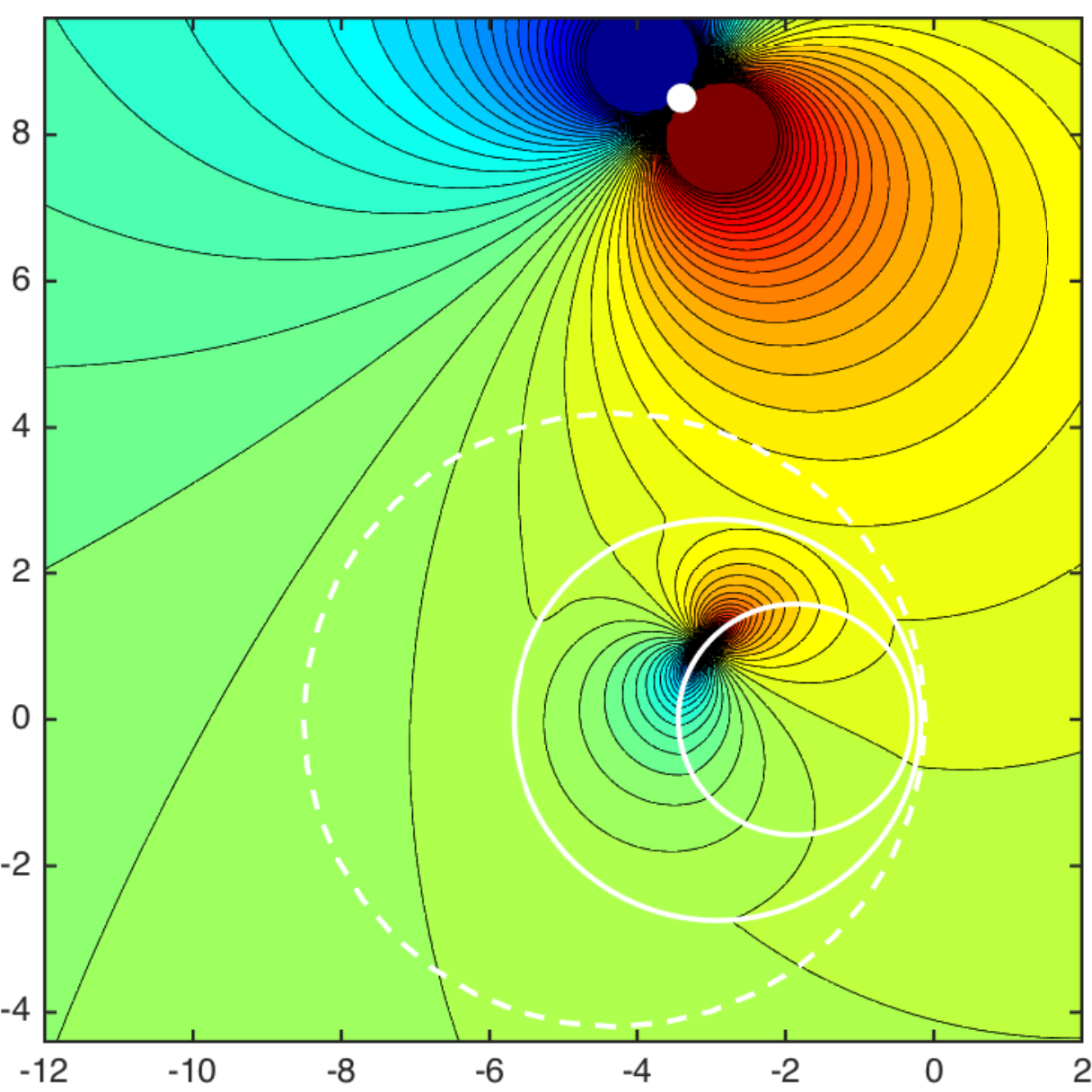,width=7cm}}\hskip0.7cm
\epsfig{figure=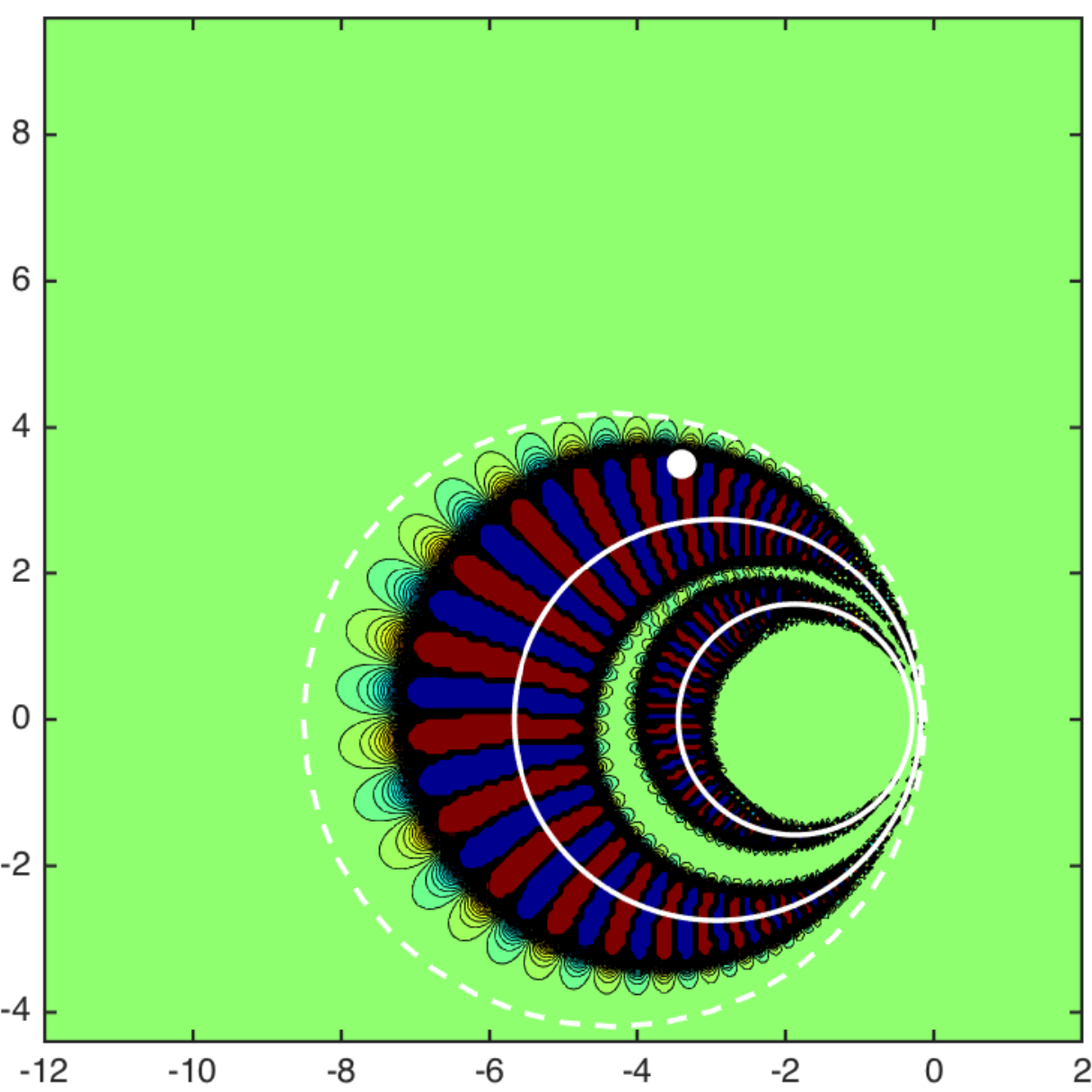,width=7cm} 
\end{center}
\caption{
{\color{black}
Cloaking for a dipole source. We set $\rho_i=0.55$, $\rho_e=0.7$ and $a=1$. The dipole source is located at $\zeta_0=(-3.4,8.5)$ in the left figure and at $\zeta_0=(-3.4,3.5)$ in the right figure. \textbf{(left)} A dipole source (small solid disk in white) is located outside the critical region $\widetilde{\Omega}_*$ (white dashed circle). The field  outside $\widetilde{\Omega}_*$ is significantly perturbed by the source. \textbf{(right)} A dipole source is located inside the critical region $\widetilde{\Omega}_*$. The anomalous resonance happens near the superlens but the field outside $\widetilde{\Omega}_*$ becomes nearly zero. The source becomes almost cloaked.
}
{The plot range is from $-10$ (blue) to $10$ (red).}
}
\label{fig_cloak_dipole}
\end{figure*}

We first present an eccentric annulus which acts as a cloaking device (Figure \ref{fig_cloak_dipole}).
{\color{black}Since we want to make a `cloaking' device, we need $\rho_b$ to satisfy the condition $\rho_b<1$. 
Setting $\rho_i=0.55$ for this example, we have $\rho_b=\rho_e^2/\rho_i=0.89<1$ ($\rho_*=\sqrt{\rho_e^3/\rho_i}=0.79$).
Applying then the M\"{o}bius transformation $\Phi$, the concentric annulus is transformed to the following eccentric structure from \eqnref{circ_to_circ}: the outer region $\widetilde{\Omega}_e=\Phi({\Omega_e})$ is the circular disk of radius $2.75$ centered at $(-2.92,0)$ and the core $\widetilde{\Omega}_i=\Phi({\Omega_i})$ is of radius $1.58$ centered at $(-1.87,0)$. 
The boundaries of the physical regions $\partial\widetilde{\Om}_i$ and $\partial\widetilde{\Om}_e$ are plotted as solid white curves in Figure \ref{fig_cloak_dipole}. 
On the other hand, the critical region's boundary $\partial\widetilde{\Omega}_*$, which is not a material interface and is the circle of radius $4.08$ centered at $(-4.55,0)$, is plotted as a dashed white circle.
We don't plot the calm region's boundary $\partial\widetilde{\Omega}_b$ in the figure for the sake of the simplicity; it is relatively close to $\p\widetilde{\Omega}_*$. 
Note that the calm region $\widetilde{\Omega}_b$ is an unbounded region whose boundary is slightly outside of $\partial\widetilde{\Omega}_*$.}

In Figure \ref{fig_cloak_dipole}(a), we assume that a dipole source {\color{black}$\widetilde F(\zeta)=\Re\{\overline{b}(\zeta-\zeta_0)^{-1}\}$ is located at $\zeta_0=(-3.4,8.5)$} with the dipole moment $b=(3,-3)$. 
The point source is plotted as a small solid disk (in white). It is clearly seen that the field distribution is smooth over the entire region {\color{black} except at the dipole source}. That is, the anomalous resonance does not happen. We can detect the dipole source by measuring the perturbation of the electric field. 

In Figure \ref{fig_cloak_dipole}(b), we change the location of the source to {\color{black}$\zeta_0 = (-3.4,3.5)$ }so that the source's location belongs to the critical region $\widetilde{\Omega}_{*}$. Then the anomalous resonance does happen as shown in the figure. As a result, the potential ouside the white dashed circle becomes nearly constant. In other words, the dipole source is almost cloaked.

\subsection{Shielding at a distance for a dipole source}

\begin{figure*}
\begin{center}
{\epsfig{figure=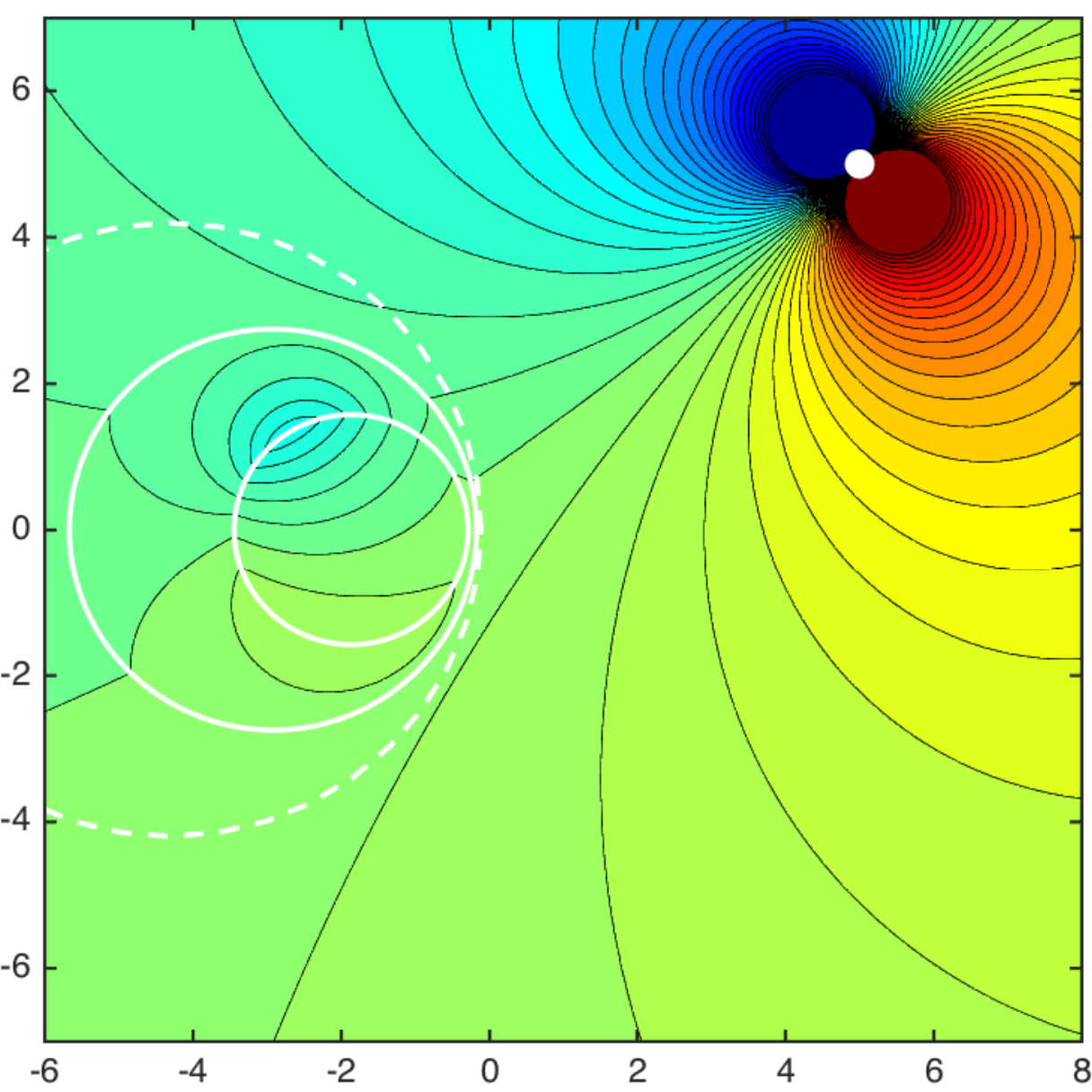,width=7cm}}\hskip0.7cm
\epsfig{figure=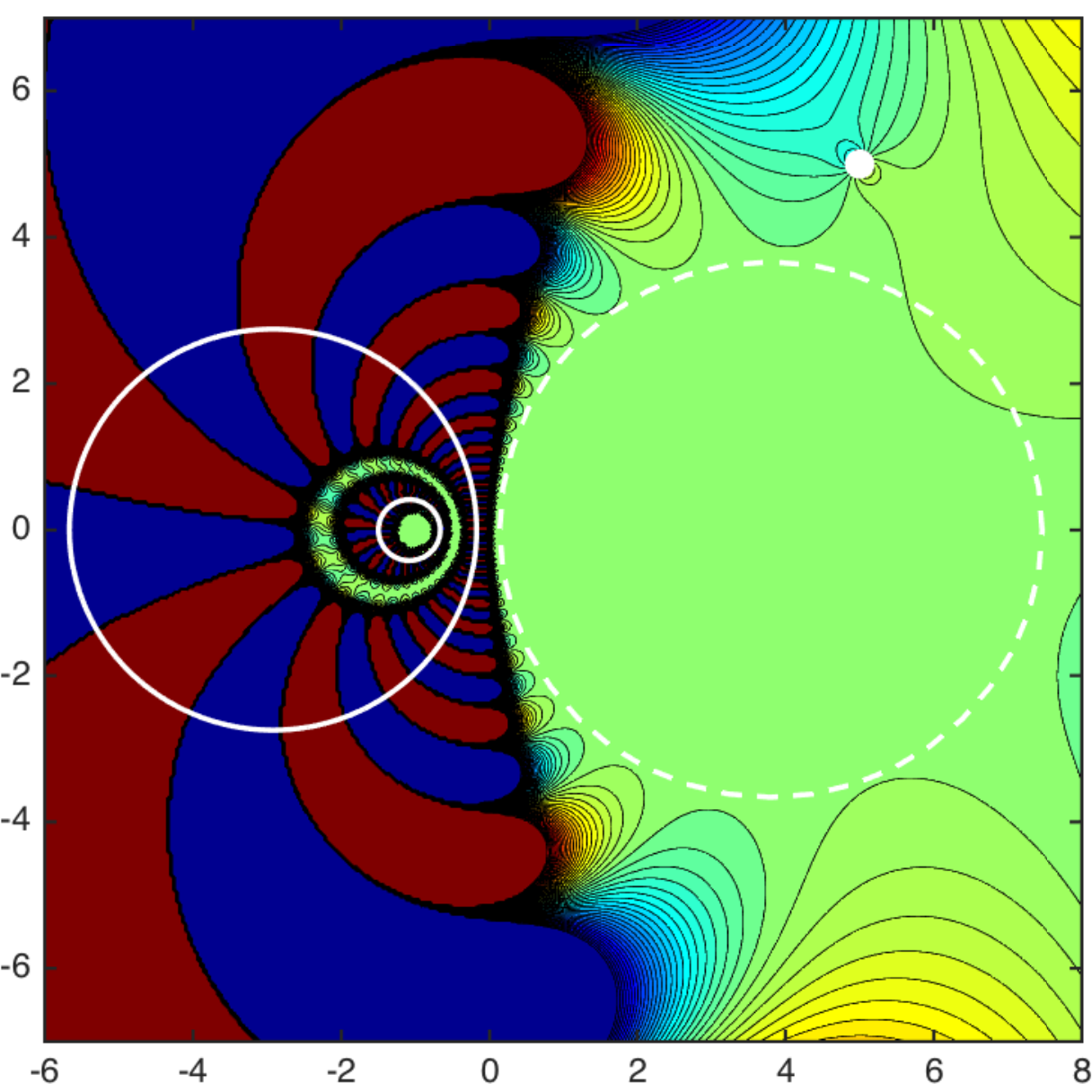,width=7cm} 
\end{center}
\caption{
{\color{black}
Shielding at a distance for a dipole source. We set $\rho_e=0.7$, $a=1$ and $\zeta_0=(5,5)$. We also set $\rho_i = 0.55$ in the left figure and $\rho_i=0.2$ in the right figure.
{\textbf{(left)}} The critical region $\widetilde{\Omega}_*$ (white dashed line) contains the eccentric superlens (white solid lines). The field  outside the white dashed circle is significantly perturbed by the source. {\textbf{(right)}} The critical region is now the region outside the white dashed circle which does not contain the superlens any longer. The field inside the white dashed circle is nearly zero and so the shielding occurs.} {The plot range is from $-10$ (blue) to $10$ (red).}
}
\label{fig_shield_dipole}
\end{figure*}

Next we show that changing the size of the core can allow shielding at a distance to happen for a dipole source (Figure \ref{fig_shield_dipole}). 

{\color{black}
In Figure \ref{fig_shield_dipole}(a), we let $\rho_i= 0.55$ as in section \ref{subsection:cloaking}. 
We also assume that a dipole source $\widetilde F(\zeta)=\Re\{\overline{b}(\zeta-\zeta_0)^{-1}\}$ is located at $\zeta_0=(5,5)$ with the dipole moment $b=(3,3)$. Since the source is located outside the critical region, the anomalous resonance does not happen. }

Now let us change the size of the core. To make the shielding at distance happen, the critical radius $\rho_*$ satisfies the condition $\rho_* > 1$. We set $\rho_i=0.2$ so that $\rho_*=\sqrt{\rho_e^3/\rho_i}=1.31>1$. Then, the core $\widetilde{\Omega}_i=\Phi({\Omega_i})$ becomes the circular disk of radius  0.42 centered at $(-1.08,0)$.
The critical region $\widetilde{\Omega}_{*}$ becomes the region outside the circle of radius $3.53$ centered at $(4.06,0)$.
The resulting eccentric annulus and the critical region are illustrated in Figure \ref{fig_shield_dipole}(b).
Note that the source is contained in the new critical region $\widetilde{\Omega}_{*}$ and $\rho_*>1$. In other words, the condition  \eqnref{shield_criteria} for shielding at a distance is satisfied. Indeed, inside the white dashed circle, the potential becomes nearly constant while there is an anomalous resonance outside.  Thus, the sheiding at a distance does happen.

\subsection{Shielding at a distance for a uniform field}
\begin{figure*}
\begin{center}
{\epsfig{figure=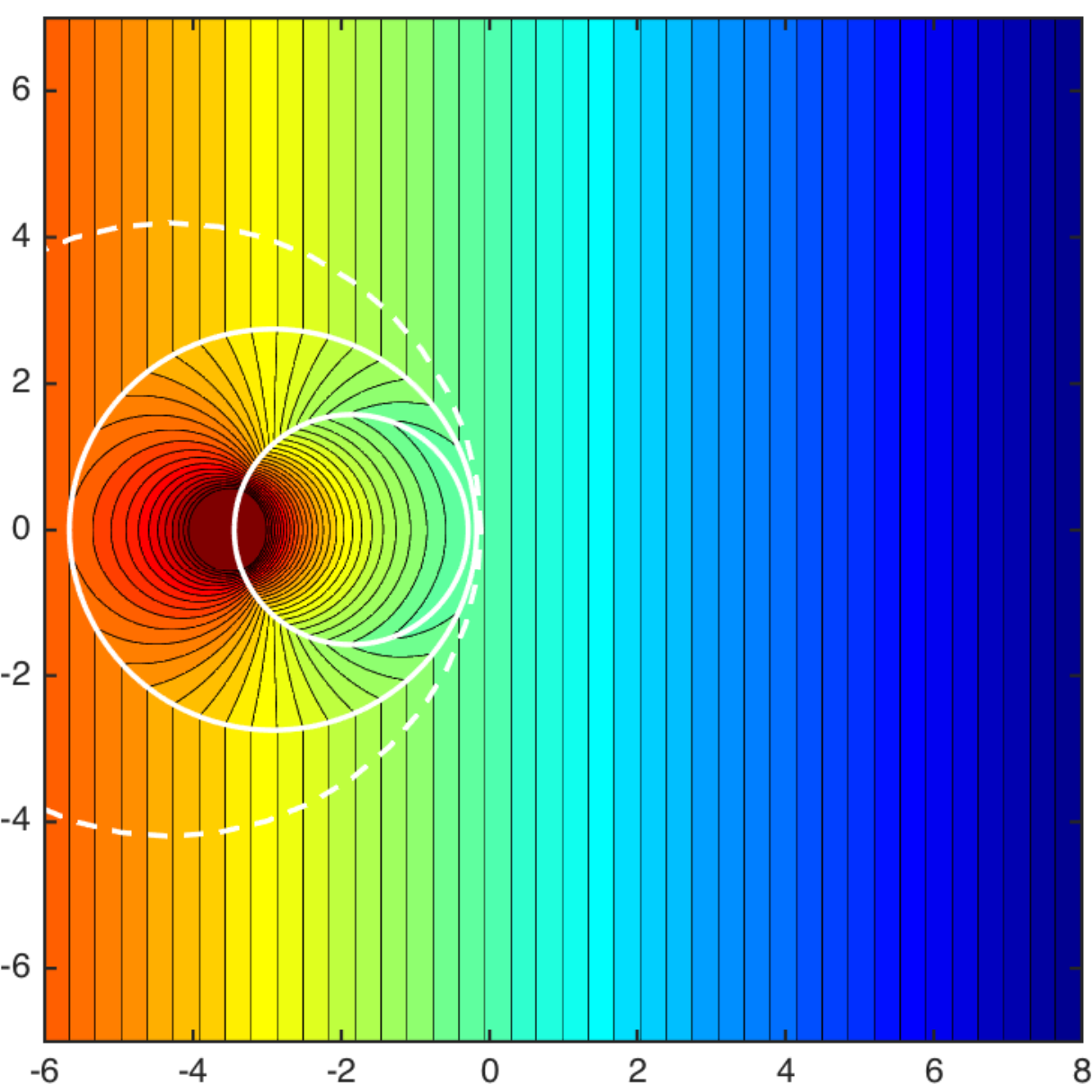,width=7cm}}\hskip0.7cm
\epsfig{figure=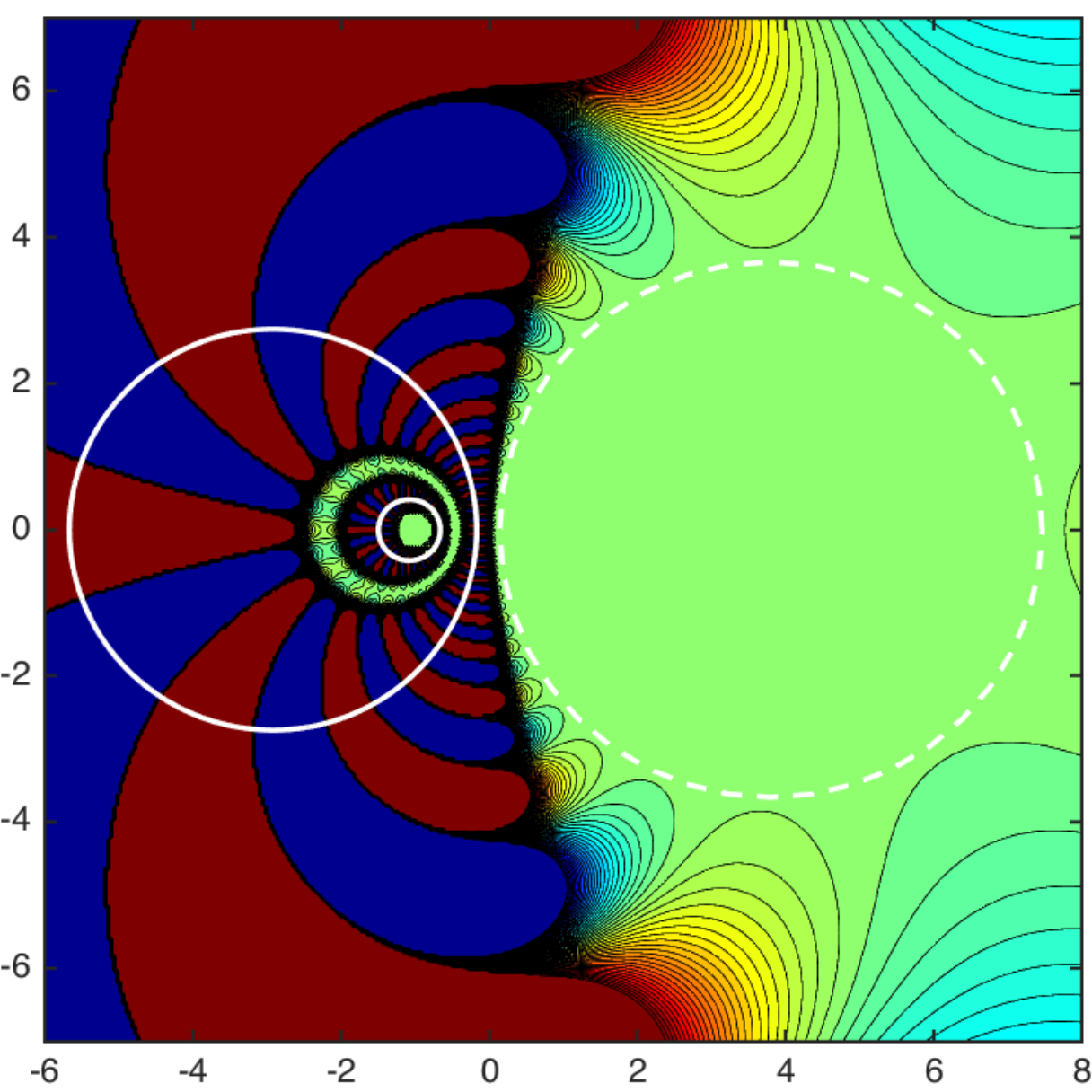,width=7cm} 
\end{center}
\caption{
Shielding at a distance for a uniform field. We set $\rho_e=0.7$ and $a=1$. We also set $\rho_i = 0.55$ in the left figure and $\rho_i=0.2$ in the right figure.
{\textbf{(left)}} The critical region $\widetilde{\Omega}_*$ (white dashed line) contains the eccentric superlens (white solid lines). {\color{black}The uniform incident field is nearly unperturbed  outside the white dashed circle.} {\textbf{(right)}} The critical region $\widetilde{\Omega}_*$ is the region outside the white dashed circle which does not contain the superlens any longer. The field inside the white dashed circle is nearly zero and so the shielding occurs. {The plot range is from $-15$ (blue) to $15$ (red).}
}

\label{fig_shield_uniform}
\end{figure*}

Finally, we consider shielding at a distance for a uniform field (Figure \ref{fig_shield_uniform}). We keep the parameters $a,\rho_i$ and $\rho_e$ as in the previous example but change the dipole source to a uniform field $\widetilde F(\zeta)=-\Re\{E_0 \zeta\}$ with $E_0=1$. As mentioned previously, an external  field can be considered as a point source located at infinity. 

In the left figure, the critical region does not contain infinity. So the anomalous resonance does not happen. The uniform field can be easily detected. In the right figure, we changed the core as in the previous example. Now the critical region (the region outside the white dashed circle) does contain infinity. So the anomalous resonance does happen. Again, the potential becomes nearly constant in the region inside the dashed circle. It means there is shielding at a distance for a uniform field.

\section{Conclusion}
We considered an eccentric superlens and showed that it exhibits a new kind of shielding effect. 
In contrast to the conventional shielding device, the new shielding effect does not require any material which encloses the region to be shielded.
In this paper, we assumed a quasistatic regime in two dimensional space. 
{
In our approach, the M\"{o}bius map $\Phi$ is essentially used and it can be represented in terms of the bipolar coordinates.
We expect that our result can be extended to the three-dimensional quasistatic case by using the bispherical coordinates.
It would also be interesting to 
extend our result
 for general electromagnetic waves by considering the full Maxwell's equations.
}



\begin{thebibliography}{10}

\bibitem{ACKLM13}
H. Ammari, G. Ciraolo, H. Kang, H. Lee, and G. W. Milton,
\newblock Spectral theory of a Neumann-Poincar\'{e}-type operator and analysis of cloaking due to anomalous localized resonance,
\newblock {\em Archive for Rational Mechanics and Analysis} 208 (2013), 667--692. 


\bibitem{ACKLM13_PRSLS}
H. Ammari, G. Ciraolo, H. Kang, H. Lee, and G. W. Milton, 
\newblock Anomalous localized resonance using a folded geometry in three dimensions,
\newblock {\em Proc. R. Soc. Lond. Ser. A} 469 (2013), 20130048.


\bibitem{Nguyen15}
 H.-M. Nguyen,
 \newblock Cloaking via anomalous localized resonance. A connection between the localized
resonance and the blow up of the power for doubly complementary media, 
\newblock {\em C. R. Math. Acad. Sci. Paris} 353 (2015), 41--46.

\bibitem{Nguyen15_JEMS} 
H.-M. Nguyen, 
\newblock Cloaking via anomalous localized resonance for doubly complementary media
in the quasi static regime, 
\newblock {\em J. Eur. Math. Soc. (JEMS)} 17 (2015), 1327--1365.


\bibitem{KSW14_CMP}
 R. V. Kohn, J. Lu, B. Schweizer and M. I. Weinstein,
 \newblock A variational perspective on cloaking by anomalous localized resonance,
 \newblock {\em Communications in Mathematical Physics}, 2014, 328(1), 1--27.




\bibitem{MNMP_PRSA_05} {G. W. Milton, N. A. Nicorovici, R. C. McPhedran,
and V. A. Podolskiy}, {A proof of superlensing in the quasistatic
regime, and limitations of superlenses in this regime due to
anomalous localized resonance}, {\em Proc. R. Soc. A} 461 (2005),
3999--4034.




\bibitem{NMM94}
N. A. Nicorovici, R. R. McPhedran, G. W. Milton,
\newblock Optical and dielectric properties of partially resonant composites,
\newblock {\em Physical Review B}, 1994, 49(12), 8479.

\bibitem{MN06}
G. W. Milton and N.-A.P. Nicorovici,
\newblock On the cloaking effects associated with anomalous localized resonance,
\newblock {\em Proceedings of the Royal Society A: Mathematical, Physical and Engineering Science}, 462 (2006), 3027--3059.



\bibitem{Pendry00}
J. B. Pendry,
\newblock Negative refraction makes a perfect lens,
\newblock {\em Phys. Rev. Lett.}, 85 (2000), 3966--3969.

\bibitem{Pendry03}
J. B. Pendry, 
\newblock Perfect cylindrical lenses,
\newblock {\em Optics Express}, 11 (2003), 755--760.

\bibitem{Ramakrishna05} S. A. Ramakrishna, 
\newblock Physics of negative refractive index materials,
\newblock {\em Rep. Prog. Phys.}, 68 (2005), 449--521. 




\bibitem{Veselago} V. G. Veselago,
\newblock The electrodynamics of substances with simultaneously negative values of $\epsilon$ and $\mu$,
\newblock {\em Usp. Fiz. Nauk}, 92 (1964), 517--526.




\end{thebibliography}
\end{document}